\documentclass[11pt]{revtex4}    
\usepackage{amssymb,epsf}
\usepackage{latexsym}
\begin{document}

\title{ Magnetic Branes in $(n+1)$-dimensional Einstein-Maxwell-dilaton gravity}
\author{A.
Sheykhi$^{1}$ \footnote{email: asheykhi@mail.uk.ac.ir}, M. H.
Dehghani$^{1,2}$ \footnote{email: mhd@shirazu.ac.ir} and  N.
Riazi$^{1}$ \footnote{email: riazi@physics.susc.ac.ir}}
\affiliation{$1$. Physics Department and Biruni Observatory, Shiraz University, Shiraz 71454, Iran\\
         $2$. Research Institute for Astrophysics and Astronomy of Maragha (RIAAM), Maragha, Iran}

\begin{abstract}
We construct two new classes of spacetimes generated by spinning
and traveling magnetic sources in
$(n+1)$-dimensional Einstein-Maxwell-dilaton gravity with
Liouville-type potential. These solutions are neither
asymptotically flat nor (A)dS. The first class of solutions which yields
a $(n+1)$-dimensional spacetime with a longitudinal magnetic field
and $k$ rotation parameters have no
curvature singularity and no horizons, but have a conic geometry.
We show that when one or more of the rotation parameters are nonzero,
the spinning branes has a net electric charge that is
proportional to the magnitude of the rotation parameters. The
second class of solutions yields a static spacetime with an angular magnetic field, and
have no curvature singularity, no horizons, and no conical singularity. Although one
may add linear momentum to the second class of solutions by a boost transformation,
one does not obtain a new solution. We find that
the net electric charge of these traveling branes with one or more
nonzero boost parameters is proportional to the magnitude of the
velocity of the branes. We also use the counterterm method
and calculate the conserved quantities of the solutions.

\end{abstract}
\pacs{04.20.Jb, 04.40.Nr, 04.50.+h}
\maketitle
\section{Introduction}
In many unified theories, including string theory, dilatons
appear. The appearance of dilaton changes the asymptotic behavior
of the solutions to be neither asymptotically flat nor (anti)-de
Sitter [(A)dS]. A motivation to investigate nonasymptotically
flat, nonasymptotically AdS solutions of Einstein gravity is that
these might lead to possible extensions of AdS/CFT correspondence.
Indeed, it has been speculated that the linear dilaton spacetimes,
which arise as near-horizon limits of dilatonic black holes, might
exhibit holography \cite{Ahar}. Another motivation is that such
solutions may be used to extend the range of validity of methods
and tools originally developed for, and tested in the case of,
asymptotically flat or asymptotically AdS black holes.
Specifically, we will find that the counterterm method inspired by
the AdS/CFT correspondence may be applied successfully to the
computation of the conserved quantities of nonasymptotically AdS
rotating magnetic branes with flat boundary at constant $r$ and
$t$. Exact dilaton black hole solutions in the absence of dilaton
potential have been constructed by many authors \cite{CDB1,
CDB2,MW,PW}. In the presence of Liouville-type potential, static
charged black hole solutions have been discovered with positive
\cite{CHM}, zero or negative constant curvature horizons
\cite{Cai}. Recently, properties of these black hole solutions
which are not asymptotically AdS or dS, have been studied
\cite{Clem}. Also, exact spherically symmetric dyonic black hole
solutions in four-dimensional and higher dimensional
Einstein-Maxwell-dilaton gravity with Liouville- type potentials
have been considered \cite{Yaz}. These exact solutions mentioned
above \cite{CDB1, CDB2,MW,PW,CHM,Cai,Clem,Yaz} are all static.
Exact rotating solutions to the Einstein equation coupled to
matter fields with curved horizons are difficult to find except in
a limited number of cases. Indeed, rotating solutions of EMd
gravity with curved horizons have been obtained only for some
limited values of the coupling constant \cite{Fr,kun,Har,kunz}.
For general dilaton coupling, the properties of rotating charged
dilaton black holes only with infinitesimally small charge
\cite{Cas} or small angular momentum in four \cite{Hor2} and five
\cite{SR} dimensions have been investigated. For arbitrary values
of angular momentum and charge, only a numerical investigation has
been done \cite{Klei}. When the horizons are flat, charged
rotating dilaton black string solutions, in four-dimensional EMd
gravity have been constructed \cite{Deh1}. Recently, these
solutions have been generalized to the $(n+1)$-dimensional EMd
gravity for an arbitrary dilaton coupling and $k$ rotation
parameters \cite{SDRP}.

In this paper we want to construct $(n+1)$-dimensional horizonless
solutions of EMd gravity. The motivation for constructing these
kinds of solutions is that they may be interpretted as cosmic
strings. Cosmic strings are topological defects that arise from
the possible phase transitions in the early universe, and may play
an important role in the formation of primordial structures. There
are many papers which are dealing directly with the issue of
spacetimes in the context of cosmic string theory \cite{Vil}. All
of these solutions are horizonless and have a conical geometry;
they are everywhere flat except at the location of the line
source. The spacetime can be obtained from the flat spacetime by
cutting out a wedge and identifying its edges. An extension to
include the electromagnetic field has also been done \cite{Muk}.
Asymptotically AdS spacetimes generated by static and spinning
magnetic sources in three and four dimensional Einstein-Maxwell
gravity with negative cosmological constant have been investigated
in \cite {Lem1,Lem2}. The generalization of these asymptotically
AdS magnetic rotating solutions of the Einstein-Maxwell equation
to higher dimensions \cite{Deh2} and higher derivative gravity
\cite{Deh3} have been also done. In the context of electromagnetic
cosmic string, it has been shown that there are cosmic strings,
known as superconducting cosmic strings, that behave as
superconductors and have interesting interactions with
astrophysical magnetic fields \cite{Wit}. The properties of these
superconducting cosmic strings have been investigated in
\cite{Moss}. Superconducting cosmic strings have also been studied
in Brans-Dicke theory \cite{Sen1}, and in dilaton gravity
\cite{Fer}. In EMd gravity, the exact magnetic rotating solutions
in three dimensions have been presented in \cite {Dia}, while two
classes of magnetic rotating solutions in four-dimensional EMd
gravity with Liouville-type potential have been constructed by one
of us \cite{Deh4}. These solutions \cite{Dia,Deh4} are not black
holes, and represent spacetimes with conic singularities. The
possibility that spacetime may have more than four dimensions is
now a standard assumption in high energy physics. Thus, it is
worth to generalize the 4-dimensional solutions of Ref.
\cite{Deh4}  to \ the case of $(n+1)$-dimensional EMd gravity for
an arbitrary value of coupling constant and investigate their
properties.

The organization of our paper is as follows: In Sec.\ref{field} we have a
brief review of the field equations and general formalism of calculating the
conserved quantities. In Sec \ref{mag}, we present the $(n+1)$-dimensional
magnetic solutions of EMd gravity with longitudinal and angular magnetic
fields, and investigate their properties. We finish our paper with some
concluding remarks.

\section{Field Equations and Conserved Quantities\label{field}}

The action of Einstein-Maxwell dilaton gravity with one scalar field $\Phi $
with Liouville-type potential in $(n+1)$ dimensions can be written as \cite
{CHM}
\begin{eqnarray}
I_{G} &=&-\frac{1}{16\pi }\int_{\mathcal{M}}d^{n+1}x\sqrt{-g}\left( \mathcal{%
R}\text{ }-\frac{4}{n-1}(\nabla \Phi )^{2}-2\Lambda e^{4\alpha \Phi
/(n-1)}-e^{-4\alpha \Phi /(n-1)}F^{2}\right)   \nonumber \\
&&-\frac{1}{8\pi }\int_{\partial \mathcal{M}}d^{n}x\sqrt{-\gamma }\Theta
(\gamma ),  \label{Act}
\end{eqnarray}
where $\mathcal{R}$ is the Ricci scalar curvature, $\Phi $ is the
dilaton field and $\Lambda $ is a constant which may be referred
to as the cosmological constant, since in the absence of the
dilaton field ($\Phi =0$) the action (\ref{Act}) reduces to the
action of Einstein-Maxwell gravity with cosmological constant.
$\alpha $ is a constant determining the strength of coupling of
the scalar and electromagnetic fields, $F^{2}=F_{\mu \nu }F^{\mu
\nu }$, where $F_{\mu \nu }=\partial _{\mu }A_{\nu }-\partial
_{\nu }A_{\mu }$ is the electromagnetic tensor field and $A_{\mu
}$ is the vector potential. The manifold $\mathcal{M}$ has metric
$g_{\mu \nu }$ and covariant derivative $\nabla _{\mu }$. $\Theta
$ is the trace of the extrinsic curvature $\Theta ^{ab}$ of any
boundary(ies) $\partial \mathcal{M}$ of the manifold
$\mathcal{M}$, with induced metric $\gamma _{ij}$. The first
integral of Eq. (\ref{Act}) does not have a well-defined
variational principle, since one encounters a total
derivative that produces a surface integral involving the derivative of $%
\delta g_{\mu \nu }$ normal to the boundary. These normal derivative terms
do not vanish by themselves, but as in the case of Einstein gravity, they
are canceled by the variation of the Gibbons-Hawking surface term (second integral).

The equations of
motion can be obtained by varying the action (\ref{Act}) with respect to the
gravitational field $g_{\mu \nu }$, the dilaton field $\Phi $ and the gauge
field $A_{\mu }$ which yields the following field equations
\begin{equation}
\mathcal{R}_{\mu \nu }=\frac{4}{n-1}\left( \partial _{\mu }\Phi \partial
_{\nu }\Phi +\frac{\Lambda }{2}e^{4\alpha \Phi /(n-1)}g_{\mu \nu }\right)
+2e^{-4\alpha \Phi /(n-1)}\left( F_{\mu \eta }F_{\nu }^{\text{ }\eta }-\frac{%
g_{\mu \nu }}{2(n-1)}F^{2}\right) ,  \label{FE1}
\end{equation}
\begin{equation}
\nabla ^{2}\Phi =\Lambda \alpha e^{4\alpha \Phi /(n-1)}-\frac{\alpha }{2}%
e^{-4\alpha \Phi /(n-1)}F^{2},  \label{FE2}
\end{equation}
\begin{equation}
\nabla _{\mu }\left( e^{-4\alpha \Phi /(n-1)}F^{\mu \nu }\right) =0.
\label{FE3}
\end{equation}

The conserved mass and angular momentum of the solutions of the above field
equations can be calculated through the use of the substraction method of
Brown and York \cite{BY}. Such a procedure causes the resulting physical
quantities to depend on the choice of reference background. For
asymptotically (A)dS solutions, the way that one deals with these
divergences is through the use of counterterm method inspired by (A)dS/CFT
correspondence \cite{Mal}. However, in the presence of a non-trivial dilaton
field, the spacetime may not behave as either dS ($\Lambda >0$) or AdS ($%
\Lambda <0$). In fact, it has been shown that with the exception of a pure
cosmological constant potential, where $\alpha =0$, no AdS or dS static
spherically symmetric solution exist for Liouville-type potential \cite{PW}.
But, as in the case of asymptotically AdS spacetimes, according to the
domain-wall/QFT (quantum field theory) correspondence \cite{Sken}, there may
be a suitable counterterm for the stress energy tensor which removes the
divergences. In this paper, we deal with spacetimes with zero curvature
boundary [$R_{abcd}(\gamma )=0$], and therefore the counterterm for the
stress energy tensor should be proportional to $\gamma ^{ab}$. Thus, the
finite stress-energy tensor in $(n+1)$-dimensional Einstein-dilaton gravity
with Liouville-type potential may be written as
\begin{equation}
T^{ab}=\frac{1}{8\pi }\left[ \Theta ^{ab}-\Theta \gamma ^{ab}+\frac{n-1}{l_{%
\mathrm{eff}}}\gamma ^{ab}\right] ,  \label{Stres}
\end{equation}
where $l_{\mathrm{eff}}$ is given by
\begin{equation}
l_{\mathrm{eff}}^{2}=\frac{(n-1)(\alpha ^{2}-n)}{2\Lambda }e^{-4\alpha \Phi
/(n-1)}.  \label{leff}
\end{equation}
In the particular case $\alpha =0$, the effective
$l_{\mathrm{eff}}^{2}$ of Eq. (\ref{leff}) reduces to
$l^{2}=-n(n-1)/2\Lambda $ of the AdS spacetimes. The first two
terms in Eq. (\ref{Stres}) are the variation of the action (\ref
{Act}) with respect to $\gamma _{ab}$, and the last term is the
counterterm which removes the divergences. One may note that the
counterterm has the same form as in the case of asymptotically AdS
solutions with zero curvature boundary, where $l$ is replaced by
$l_{\mathrm{eff}}$. To compute the
conserved charges of the spacetime, one should choose a spacelike surface $%
\mathcal{B}$ in $\partial \mathcal{M}$ with metric $\sigma _{ij}$, and write
the boundary metric in ADM (Arnowitt-Deser-Misner) form:
\[
\gamma _{ab}dx^{a}dx^{a}=-N^{2}dt^{2}+\sigma _{ij}\left( d\varphi
^{i}+V^{i}dt\right) \left( d\varphi ^{j}+V^{j}dt\right) ,
\]
where the coordinates $\varphi ^{i}$ are the angular variables
parameterizing the hypersurface of constant $r$ around the origin, and $N$
and $V^{i}$ are the lapse and shift functions, respectively. When there is a
Killing vector field $\mathcal{\xi }$ on the boundary, then the quasilocal
conserved quantities associated with the stress tensors of Eq. (\ref{Stres})
can be written as
\begin{equation}
Q(\mathcal{\xi )}=\int_{\mathcal{B}}d^{n-1} x \sqrt{\sigma }T_{ab}n^{a}%
\mathcal{\xi }^{b},  \label{charge}
\end{equation}
where $\sigma $ is the determinant of the metric $\sigma _{ij}$, $\mathcal{%
\xi } $ and $n^{a}$ are the Killing vector field and the unit normal vector
on the boundary $\mathcal{B}$. For boundaries with timelike ($\xi =\partial
/\partial t $), rotational ($\varsigma_{i} =\partial /\partial \phi^{i} $)
and translational Killing vector fields ($\zeta_{i} =\partial /\partial x^{i}
$), one obtains the quasilocal mass, components of total angular and linear
momenta as
\begin{eqnarray}
M &=&\int_{\mathcal{B}}d^{n-1} x \sqrt{\sigma }T_{ab}n^{a}\xi ^{b},
\label{Mastot} \\
J_{i} &=&\int_{\mathcal{B}}d^{n-1} x \sqrt{\sigma }T_{ab}n^{a}\varsigma_{i}
^{b},  \label{Angtot} \\
P_{i} &=&\int_{\mathcal{B}}d^{n-1} x \sqrt{\sigma }T_{ab}n^{a}\zeta_{i} ^{b},
\label{Lintot}
\end{eqnarray}
provided the surface $\mathcal{B}$ contains the orbits of $\varsigma $.
These quantities are, respectively, the conserved mass, angular and linear
momenta of the system enclosed by the boundary $\mathcal{B}$. Note that they
will both be dependent on the location of the boundary $\mathcal{B}$ in the
spacetime, although each is independent of the particular choice of
foliation $\mathcal{B}$ within the surface $\partial \mathcal{M}$.

\section{(n+1)-dimensional Magnetic Rotating Solutions \label{mag}}

In this section, we obtain the $(n+1)$-dimensional horizonless solutions of
Eqs. (\ref{FE1})-(\ref{FE3}). First, we construct a spacetime generated by a
magnetic source which produces a longitudinal magnetic field. Second, we
obtain a spacetime generated by a magnetic source that produces angular
magnetic fields along the $\phi _{i}$ coordinates.

\subsection{Longitudinal magnetic field solutions\label{Lon1}}

Here we want to obtain the $(n+1)$-dimensional solutions of Eqs. (\ref{FE1}%
)-(\ref{FE3}) which produce longitudinal magnetic fields in the Euclidean
submanifold spanned by the $x^{i}$ coordinates ($i=1,...,n-2$). We assume that
the metric has the following form:
\begin{equation}
ds^{2}=-\frac{\rho ^{2}}{l^{2}}R^{2}(\rho )dt^{2}+\frac{d\rho ^{2}}{f(\rho )}%
+l^{2}f(\rho )d\phi ^{2}+\frac{\rho ^{2}}{l^{2}}R^{2}(\rho
)dX^{2}{,} \label{Met1}
\end{equation}
where $dX^{2}={{\sum_{i=1}^{n-2}}}(dx^{i})^{2}$. Note that the coordinates $%
x^{i}$ have the dimension of length, while the angular coordinate $\phi $ is
dimensionless as usual and ranges in $0\leq \phi <2\pi $. The motivation for
this metric gauge $[g_{tt}\varpropto -\rho ^{2}$ and $(g_{\rho \rho
})^{-1}\varpropto g_{\phi \phi }]$ instead of the usual Schwarzschild gauge $%
[(g_{\rho \rho })^{-1}\varpropto g_{tt}$ and $g_{\phi \phi }\varpropto \rho
^{2}]$ comes from the fact that we are looking for a string solution with
conic singularity. The Maxwell equation (\ref{FE3}) can be integrated
immediately to give
\begin{equation}  \label{Ftr}
F_{\phi \rho}=\frac{ql e^{4\alpha \Phi /(n-1)}}{(\rho R)^{n-1}}
\end{equation}
where $q$, is an integration constant related to the electric charge of the
brane. In order to solve the system of equations (\ref{FE1}) and (\ref{FE2})
for three unknown functions $f(\rho )$, $R(\rho )$ and $\Phi (\rho )$, we
make the ansatz
\begin{equation}  \label{Rphi}
R(\rho )=e^{2\alpha \Phi /(n-1)}.
\end{equation}
Using (\ref{Rphi}), the Maxwell fields (\ref{Ftr}) and the metric (\ref{Met1}%
), one can easily show that Eqs. (\ref{FE1}) and (\ref{FE2}) may be written
as
\begin{eqnarray}
&&f\Phi^{\prime \prime }+f^{\prime }\Phi^{\prime }+2\alpha f {\Phi^{\prime }}%
^2+(n-1)\rho^{-1}f\Phi^{\prime }-\alpha \Lambda e^{4\alpha
\Phi/(n-1)} \nonumber  \\
&&+\alpha q^2 \rho^{2-2n}e^{-4\alpha \Phi(n-2)/(n-1)}=0,  \label{Eqphi} \\
&& (n-1)f^{\prime \prime }+(n-1)^2\rho^{-1}f^{\prime}+2(n-1)\alpha
f^{\prime }\Phi^{\prime }+4\Lambda e^{4\alpha \Phi/(n-1)} \nonumber \\
&& +4(n-2) q^2 \rho^{2-2n}e^{-4\alpha \Phi(n-2)/(n-1)}=0,  \label{Eqpp} \\
&& (n-1)\rho^{-1}f^{\prime}+2\alpha f^{\prime}
\phi^{\prime}+2\alpha
f\phi^{\prime \prime}+4\alpha(n-1)\rho^{-1}f \Phi^{\prime}+4\alpha^2f{%
\Phi^{\prime}}^2  \nonumber  \\
&&+(n-1)(n-2)\rho^{-2}f+2\Lambda e^{4\alpha \Phi/(n-1)}-2q^2
\rho^{2-2n}e^{-4\alpha \Phi(n-2)/(n-1)}=0,  \label{Eqtt} \\
&& (n-1)f^{\prime \prime}+(n-1)^2\rho^{-1}
f^{\prime}+2\alpha(n-1)f^{\prime} \phi^{\prime}+4\alpha (n-1)f
\phi^{\prime \prime}+8\alpha(n-1)\rho^{-1}f
\Phi^{\prime} \nonumber\\
&&+8(\alpha^2+1) f{\Phi^{\prime}}^2+4\Lambda e^{4\alpha
\Phi/(n-1)}+4(n-2)q^2 \rho^{2-2n}e^{-4\alpha \Phi(n-2)/(n-1)}=0,
\label{Eqrr}
\end{eqnarray}
where the prime denotes a derivative with respect to the $\rho $ coordinate.
The above equations (\ref{Eqphi})-(\ref{Eqrr}) are coupled differential
equations for the unknown functions $f(\rho)$ and $\Phi (\rho)$. Combining
Eqs. (\ref{Eqpp}) and (\ref{Eqrr}), one can find an uncoupled differential
equation for $\Phi (\rho)$ as
\begin{equation}
(n-1)\alpha (\rho\Phi^{\prime \prime}+2\Phi^{\prime})+2 (1+\alpha^2)\rho {%
\Phi^{\prime}}^2 =0,  \label{EqPhir}
\end{equation}
with the solution
\begin{equation}
\Phi (\rho )=\frac{(n-1)\alpha }{2(1+\alpha ^{2})}\ln (\frac{b}{\rho }),
\label{Phir}
\end{equation}
where $b$ is an arbitrary constant. Substituting $\Phi (\rho )$ given by Eq.
(\ref{Phir}) in the field equations (\ref{Eqphi})-(\ref{Eqrr}) one finds the
function $f(\rho)$ as
\begin{equation}
f(\rho )=\frac{2\Lambda (\alpha ^{2}+1)^{2}b^{2\gamma }}{(n-1)(\alpha ^{2}-n)%
}\rho ^{2(1-\gamma )}+\frac{m}{\rho ^{(n-1)(1-\gamma )-1}}-\frac{%
2q^{2}(\alpha ^{2}+1)^{2}b^{-2(n-2)\gamma }}{(n-1)(\alpha ^{2}+n-2)\rho
^{2(n-2)(1-\gamma )}},
\end{equation}
where $m$ is an arbitrary constant and $\gamma =\alpha ^{2}/(\alpha ^{2}+1)$.

In order to study the general structure of these solutions, we first look
for curvature singularities. It is easy to show that the Kretschmann scalar $%
R_{\mu \nu \lambda \kappa }R^{\mu \nu \lambda \kappa }$ diverges
at $\rho =0$ and therefore one might think that there is a
curvature singularity located at $\rho =0$. However, as we will
see below, the spacetime will never achieve $\rho =0$. The
function $f(\rho )$ is negative for $\rho <r_{+}$ and
positive for $\rho >r_{+}$, where $r_{+}$ is the largest root of $f(\rho )=0$%
. Indeed, $g_{\rho \rho }$ and $g_{\phi \phi }$ are related by $f(\rho
)=g_{\rho \rho }^{-1}=l^{-2}g_{\phi \phi }$, and therefore when $g_{\rho
\rho }$ becomes negative (which occurs for$\rho <r_{+}$) so does $g_{\phi
\phi }$. This leads to apparent change of signature of the metric from $%
(n-1)+$ to $(n-2)+$ as one extends the spacetime to $\rho <r_{+}$. This
indicates that we are using an incorrect extension. To get rid of this
incorrect extension, we introduce the new radial coordinate $r$ as
\[
r^{2}=\rho ^{2}-r_{+}^{2}\Rightarrow d\rho ^{2}=\frac{r^{2}}{r^{2}+r_{+}^{2}}%
dr^{2}.
\]
With this new coordinate, the metric (\ref{Met1}) is
\begin{eqnarray}
ds^{2}
&=&-\frac{r^{2}+r_{+}^{2}}{l^{2}}R^{2}(r)dt^{2}+l^{2}f(r)d\phi
^{2}
\nonumber \\
&&+\frac{r^{2}}{(r^{2}+r_{+}^{2})f(r)}dr^{2}+\frac{r^{2}+r_{+}^{2}}{l^{2}}%
R^{2}(r)dX^{2},  \label{metric2}
\end{eqnarray}
where the coordinates $r$ assumes the values $0\leq r<\infty $, and $f(r)$, $%
R(r)$ and $\Phi (r)$ are now given as
\begin{eqnarray}
f(r) &=&\frac{2\Lambda (\alpha ^{2}+1)^{2}b^{2\gamma }}{(n-1)(\alpha ^{2}-n)}%
(r^{2}+r_{+}^{2})^{(1-\gamma )}+\frac{m}{(r^{2}+r_{+}^{2})^{\left[
(n-1)(1-\gamma )-1\right] /2}}  \nonumber \\
&&-\frac{2q^{2}(\alpha ^{2}+1)^{2}b^{-2(n-2)\gamma }}{(n-1)(\alpha
^{2}+n-2)(r^{2}+r_{+}^{2})^{(n-2)(1-\gamma )}},  \label{f2}
\end{eqnarray}
\begin{equation}
R(r)=\frac{b^{\gamma }}{\left( r^{2}+r_{+}^{2}\right) ^{\gamma /2}}, \hspace{%
0.5cm} \Phi (r)=\frac{(n-1)\alpha }{4(1+\alpha ^{2})}\ln \left( \frac{b^{2}}{%
r^{2}+r_{+}^{2}}\right). \label{R2}
\end{equation}
The metric (\ref{metric2}) is neither
asymptotically flat nor (anti)-de Sitter. One can easily show that the Kretschmann scalar does not diverge in the
range $0\leq r<\infty $. However, the spacetime has a conic geometry and has
a conical singularity at $r=0$. In fact, using a Taylor expansion, the metric (\ref{metric2}) is written as
\begin{equation}
ds^{2} =-b^{\gamma }\frac{r_{+}^{2-\gamma }}{l^{2}}dt^{2}+\frac{G^{-1}}{%
r_{+}^{2}}dr^{2} +l^2 G r^{2}d\varphi^{2}+b^{\gamma }\frac{r_{+}^{2-\gamma }}{l^{2}}dX ^{2}, \label{metr0a}
\end{equation}
where
\begin{eqnarray}
G &=&\frac{\Lambda (\alpha ^{2}+1)^{2}b^{2\gamma }((n+1)(1-\gamma )-1)}{%
(n-1)(\alpha ^{2}-n)r_{+}^{2\gamma }}  \nonumber \\
&&+\frac{q^{2}(\alpha ^{2}+1)^{2}b^{(4-2n)\gamma }\left( (n-3)(1-\gamma
)+1\right) }{(n-1)(\alpha ^{2}+n-2)r_{+}^{2(n-2)(1-\gamma )+2}}. \label{metr0b}
\end{eqnarray}
Indeed, there is a conical singularity at $r=0$ since:
\begin{equation}
\lim_{r\rightarrow 0}\frac{1}{r}\sqrt{\frac{g_{\varphi \varphi }}{g_{rr}}}\neq 1.
\end{equation}
That is, as the radius $r$ tends to zero, the limit of the ratio ``\textrm{%
circumference/radius}'' is not $2\pi $ and therefore the spacetime has a
conical singularity at $r=0$. The canonical singularity can be removed if one identifies the coordinate $\varphi$ with the period
\begin{equation}
\textrm{Period}_{\varphi}=2 \pi \left(\lim_{r\rightarrow 0}\frac{1}{r}\sqrt{\frac{g_{\varphi \varphi }}{g_{rr}}} \right)^{-1}=2 \pi (1-4 \mu), \label{period}
\end{equation}
where $\mu$ is given by
\begin{equation}
\mu=\frac{1}{4}\left[1-\left(\frac{1}{2}\frac{ml(\alpha ^{2}+n-2)}{\alpha ^{2}+1}r_{+}^{(n-1)(\gamma -1)}+%
\frac{2(1+\alpha ^{2})}{(\alpha ^{2}-n)}\Lambda lb^{2\gamma
}r_{+}^{1-2\gamma }\right)^{-1}\right]. \label{mu}
\end{equation}
From Eqs. (\ref{metr0a})-(\ref{mu}), one concludes that near the origin $r=0$,
the metric (\ref{metric2}) describes a spacetime which is locally flat but has a conical singularity at
$r=0$ with a deficit angle $\delta \varphi=8 \pi \mu$. Since near the origin the metric (\ref{metric2})
for $n=3$ is identical to the spacetime generated by a cosmic string, by use of Vilenkin procedure \cite{Vil2},
one can show that $\mu$ of Eq. (\ref{mu})
can be interpreted as the mass per unit length of the string.

Now we investigate the casual
structure of the spacetime. As one can see from Eq. (\ref{f2}), there is no
solution for $\alpha =\sqrt{n}$ with a Liouville potential ($\Lambda \neq 0$%
). The cases with $\alpha >\sqrt{n}$ and $\alpha <\sqrt{n}$ should be
considered separately.
For $\alpha >\sqrt{n}$, as $r$ goes to infinity the dominant term in Eq. (
\ref{f2}) is the second term, and therefore the function $f(r)$ is positive
in the whole spacetime, despite the sign of the cosmological constant $%
\Lambda $, and is zero at $r=0$. Thus, the solution given by Eqs. (\ref
{metric2}) and (\ref{f2}) exhibits a spacetime with conic singularity at $%
r=0 $. For $\alpha <\sqrt{n}$, the dominant term for large values of $r$\ is
the first term, and therefore the function $f(r)$ given in Eq. (\ref{f2}) is
positive in the whole spacetime only for negative values of $\Lambda $. In
this case the solution represents a spacetime with conic singularity at $r=0$%
. The solution is not acceptable for $\alpha <\sqrt{n}$ with positive values
of $\Lambda $, since the function $f(r)$ is negative for large values of $r$%
. Of course, one may ask for the completeness of the spacetime with $r\geq 0$
(or $\rho \geq r_{+}$) \cite{Lem2,Hor3}. It is easy to see that the
spacetime described by Eq. (\ref{metric2}) is both null and timelike
geodesically complete. In fact, we can show that every null or timelike
geodesic starting from an arbitrary point can either extend to infinite
values of the affine parameter along the geodesic or end on a singularity at
$r=0$. Using the geodesic equation, one obtains
\begin{eqnarray}
&&\dot{t}=\frac{l^{2}}{b^{2\gamma }(r^{2}+r_{+}^{2})^{1-\gamma }}E,\hspace{%
0.5cm}\dot{x^{i}}=\frac{l^{2}}{b^{2\gamma }(r^{2}+r_{+}^{2})^{1-\gamma }}%
P^{i},\hspace{0.5cm}\dot{\phi}=\frac{1}{l^{2}f(r)}L,  \label{Geo1} \\
&&r^{2}\dot{r}^{2}=(r^{2}+r_{+}^{2})f(r)\left[ \frac{l^{2}(E^{2}-\mathbf{P}%
^{2})}{b^{2\gamma }(r^{2}+r_{+}^{2})^{1-\gamma }}-\eta \right] -\frac{%
r^{2}+r_{+}^{2}}{l^{2}}L^{2},  \label{Geo2}
\end{eqnarray}
where the dot denotes the derivative with respect to an affine parameter and
$\eta $ is zero for null geodesics and $+1$ for timelike geodesics. $E$, $L$%
, and $P^{i}$ are the conserved quantities associated with the coordinates $%
t $, $\phi $, and $x^{i}$ respectively, and $\mathbf{P}^{2}=%
\sum_{i=1}^{n-2}(P^{i})^{2}$. Notice that $f(r)$ is always positive for $r>0$
and zero for $r=0$.

First we consider the null geodesics ($\eta =0$). $(\mathit{i})$ If $E^{2}>%
\mathbf{P}^{2}$ the spiraling particles ($L>0$) coming from infinity have a
turning point at $r_{tp}>0$, while the nonspiraling particles ($L=0$) have a
turning point at $r_{tp}=0$. $(\mathit{ii})$ If $E=\mathbf{P}$ and $L=0$,
whatever is the value of $r$, $\dot{r}$ and $\dot{\phi}$ vanish and
therefore the null particles move in a straight line in the $(n-2)$%
-dimensional submanifold spanned by $x^{1}$ to $x^{n-2}$. $(\mathit{iii})$
For $E=\mathbf{P}$ and $L\neq 0$, and also for $E^{2}<\mathbf{P}^{2}$ and
any value of $L$, there is no possible null geodesic.

Second, we analyze the timelike geodesics ($\eta =+1$). Timelike geodesics
are possible only if $l^{2}(E^{2}-P^{2})>b^{2\gamma }r_{+}^{2(1-\gamma )}$.
In this case the turning points for the nonspiraling particles ($L=0$) are $%
r_{tp}^{1}=0$ and $r_{tp}^{2}$ given as
\begin{equation}
r_{tp}^{2}=\sqrt{[b^{-2\gamma }l^{2}(E^{2}-\mathbf{P}^{2})]^{1/(1-\gamma
)}-r_{+}^{2}},
\end{equation}
while the spiraling ($L\neq 0$) timelike particles are bound between $%
r_{tp}^{a}$ and $r_{tp}^{b}$ given by
\begin{equation}
0<r_{tp}^{a}\leq r_{tp}^{b}<r_{tp}^{2}.
\end{equation}
Therefore, we have confirmed that the spacetime described by Eq. (\ref
{metric2}) is both null and timelike geodesically complete.

\subsection{Longitudinal magnetic field solutions with all rotation
parameters\label{Lon2}}

Our aim here is to obtain the $(n+1)$-dimensional longitudinal magnetic
field solutions with a complete set of rotation parameters. The rotation
group in $n+1$ dimensions is $SO(n)$ and therefore the number of independent
rotation parameters is $[n/2]$, where $[x]$ is the integer part of $x$. We
now generalize the above metric given in Eq. (\ref{metric2}) with $k\leq
\lbrack n/2]$ rotation parameters. This generalized solution can be written
as
\begin{eqnarray}
ds^{2} &=&-\frac{r^{2}+r_{+}^{2}}{l^{2}}R^{2}(r)\left( \Xi dt-{{%
\sum_{i=1}^{k}}}a_{i}d\phi ^{i}\right) ^{2}+f(r)\left( \sqrt{\Xi ^{2}-1}dt-%
\frac{\Xi }{\sqrt{\Xi ^{2}-1}}{{\sum_{i=1}^{k}}}a_{i}d\phi ^{i}\right) ^{2}
\nonumber \\
&&+\frac{r^{2}dr^{2}}{(r^{2}+r_{+}^{2})f(r)}+\frac{r^{2}+r_{+}^{2}}{%
l^{2}(\Xi ^{2}-1)}R^{2}(r){\sum_{i<j}^{k}}(a_{i}d\phi _{j}-a_{j}d\phi
_{i})^{2}+\frac{r^{2}+r_{+}^{2}}{l^{2}}R^{2}(r)dX^{2},  \label{metric3}
\end{eqnarray}
where $\Xi =\sqrt{1+\sum_{i}^{k}a_{i}^{2}/l^{2}}$, $dX^{2}$ is the Euclidean
metric on the $(n-k-1)$-dimensional submanifold and $f(r)$ and $R(r)$ are
the same as given in Eq. (\ref{f2}). The gauge potential is
\begin{equation}
A_{\mu }=\frac{qb^{(3-n)\gamma }}{\Gamma (r^{2}+r_{+}^{2})^{\Gamma /2}}%
\left( \sqrt{\Xi ^{2}-1}\delta _{\mu }^{t}-\frac{\Xi }{\sqrt{\Xi ^{2}-1}}%
a_{i}\delta _{\mu }^{i}\right) \hspace{0.5cm}{\text{(no sum on }i\text{)}}.
\label{A3}
\end{equation}
where $\Gamma =(n-3)(1-\gamma )+1$. Again this spacetime has no horizon and
curvature singularity. However, it has a conical singularity at $r=0$. One
should note that these solutions reduce to those discussed in \cite{Deh2},
in the absence of dilaton field ($\alpha =\gamma =0$) and those presented in
\cite{Deh4} for $n=3$.

Now we calculate conserved quantities of these solutions. Denoting the
volume of the hypersurface boundary at constant $t$ and $r$ by $%
V_{n-1}=(2\pi )^{k}\Sigma _{n-k-1}$, the mass and angular momentum per unit
volume $V_{n-1}$ of the branes ($\alpha <\sqrt{n}$) can be calculated
through the use of Eqs. (\ref{Mastot}) and (\ref{Angtot}). We find
\begin{eqnarray}
{M} &=&\frac{b^{(n-1)\gamma }}{16\pi l^{n-2}}\left( \frac{(n-\alpha ^{2})\Xi
^{2}-(n-1)}{1+\alpha ^{2}}\right) m,  \label{M1} \\
J_{i} &=&\frac{b^{(n-1)\gamma }}{16\pi l^{n-2}}\left( \frac{n-\alpha ^{2}}{%
1+\alpha ^{2}}\right) \Xi ma_{i}.  \label{J1}
\end{eqnarray}
For $a_{i}=0$ ($\Xi =1$), the angular momentum per unit volume
vanishes, and therefore $a_{i}$'s are the rotational parameters of
the spacetime. Of course, one may note that in the particular case
$n=3$, these conserved charges reduce to the conserved charges of
the magnetic rotating black
string obtained in Ref. \cite{Deh4}, and in the absence of dilaton field ($%
\alpha =\gamma =0$) they reduce to those of Ref. \cite{Deh2}.

\subsection{The Angular Magnetic Field Solutions}

In subsection \ref{Lon1}, we found a spacetime generated by a magnetic
source which produces a longitudinal magnetic field along $x^{i}$
coordinates. Now, we want to obtain a spacetime generated by a magnetic
source that produces angular magnetic fields along the $\phi ^{i}$
coordinates. Following the steps of Subsection \ref{Lon1} but now with the
roles of $\phi $ and $x$ interchanged, we can directly write the metric and
vector potential satisfying the field equations (\ref{Eqphi})-(\ref{Eqrr})
as
\begin{eqnarray}
ds^{2} &=&-\frac{r^{2}+r_{+}^{2}}{l^{2}}R^{2}(r)dt^{2}+\frac{r^{2}dr^{2}}{%
(r^{2}+r_{+}^{2})f(r)}  \nonumber \\
&&+f(r)dx^{2}+(r^{2}+r_{+}^{2})R^{2}(r)d\Omega ^{2},  \label{Metr3}
\end{eqnarray}
where $d\Omega ^{2}={{\sum_{i=1}^{n-2}}}(d\phi ^{i})^{2}$ and $f(r)$ and $%
R(r)$ are given in Eq. (\ref{f2}). The angular coordinates $\phi ^{i}$'s
range in $0\leq \phi ^{i}<2\pi $. The gauge potential is now given by
\begin{equation}
A_{\mu }=\frac{qb^{(3-n)\gamma }}{\Gamma (r^{2}+r_{+}^{2})^{\Gamma /2}}%
\delta _{\mu }^{x}.  \label{Pot3}
\end{equation}
The Kretschmann scalar does not diverge for any $r$ and therefore there is
no curvature singularity. The spacetime (\ref{Metr3}) is also free of conic
singularity. In addition, it is notable to mention that the radial geodesic
passes through $r=0$ (which is free of singularity) from positive values to
negative values of the coordinate $r$. This shows that the radial coordinate
in Eq. (\ref{Metr3}) can take the values $-\infty <r<\infty $. This analysis
may suggest that one is in the presence of a traversable wormhole with a
throat of dimension $r_{+}$. However, in the vicinity of $r=0$, the metric (%
\ref{Metr3}) can be written as
\begin{equation}
ds^{2} =-b^{\gamma }\frac{r_{+}^{2-\gamma }}{l^{2}}dt^{2}+\frac{G^{-1}}{%
r_{+}^{2}}dr^{2} +Gr^{2}dx^{2}+b^{\gamma }r_{+}^{2-\gamma }d\Omega ^{2}, \label{worm}
\end{equation}
where $G$ is given in Eq. (\ref{metr0b}).
Equation (\ref{worm}) shows that, at $r=0$, the $x$ direction collapses and
therefore we have to abandon the wormhole interpretation.

To add linear momentum to the spacetime along the coordinate $x^{i}$, we
perform the boost transformation
\[
t\mapsto \Xi t-(v_{i}/l)x^{i};\text{ \ \ }x^{i}\mapsto \Xi x^{i}-(v_{i}/l)t%
\text{ \ \ (no sum on }i\text{).}
\]
in the $t-x_{i}$ plane, where $v_{i}\ $is a boost parameter and $\Xi =\sqrt{%
1+\sum_{i}^{\kappa }v_{i}^{2}/l^{2}}$ ($i$ can run from $1$ to $\kappa \leq
n-2$). One obtains
\begin{eqnarray}
ds^{2} =&&-\frac{r^{2}+r_{+}^{2}}{l^{2}}R^2(r)\left( \Xi dt-l^{-1}{{%
\sum_{i=1}^{\kappa}}}v_{i}dx^{i}\right) ^{2}+f(r)\left( \sqrt{\Xi ^{2}-1}dt-%
\frac{\Xi }{l\sqrt{\Xi ^{2}-1}}{{\sum_{i=1}^{\kappa}}}v_{i}dx^{i}\right) ^{2}
\nonumber \\
&&+\frac{r^{2}+r_{+}^{2}}{l^{4}(\Xi ^{2}-1)}R^2(r)\text{ }{%
\sum_{i<j}^{\kappa}}(v_{i}dx_{j}-v_{j}dx_{i})^{2}+\frac{r^{2}dr^{2}}{%
(r^{2}+r_{+}^{2})f(r)}+(r^{2}+r_{+}^{2})R^2(r)d\Omega^{2},  \label{Metr4}
\end{eqnarray}
The gauge potential is given by

\begin{equation}
A_{\mu }=\frac{qb^{(3-n)\gamma }}{\Gamma (r^{2}+r_{+}^{2})^{\Gamma /2}}%
\left( \sqrt{\Xi ^{2}-1}\delta _{\mu }^{t}-\frac{\Xi }{l\sqrt{\Xi ^{2}-1}}%
v_{i}\delta _{\mu }^{i}\right) \hspace{0.5cm}{\text{(no sum on i)}}.
\end{equation}
This boost transformation is permitted globally since $x^{i}$ is not an
angular coordinate. Although the boosted solution (\ref{Metr4}) generates an
electric field, it is not a new solution. The conserved quantities of the
spacetime (\ref{Metr4}) are the mass and linear momentum. The mass and
linear momentum per unit volume $V_{n-1}$ of the branes ($\alpha <\sqrt{n}$)
can be calculated through the use of Eqs. (\ref{Mastot}) and (\ref{Lintot}).
We find
\begin{equation}
{M}=\frac{b^{(n-1)\gamma }}{16\pi l^{n-2}}\left( \frac{(n-\alpha ^{2})\Xi
^{2}-(n-1)}{1+\alpha ^{2}}\right) m,
\end{equation}
\begin{equation}
P_{i}=\frac{b^{(n-1)\gamma }}{16\pi l^{n-2}}\left( \frac{n-\alpha ^{2}}{%
1+\alpha ^{2}}\right) \Xi mv_{i}.
\end{equation}

Finally, we calculate the electric charge of the solutions (\ref{metric3})
and (\ref{Metr4}) obtained in this section. To determine the electric field
we should consider the projections of the electromagnetic field tensor on
special hypersurfaces. The normal to such hypersurfaces is
\[
u^{0}=\frac{1}{N},\text{ \ }u^{r}=0,\text{ \ }u^{i}=-\frac{V^{i}}{N},
\]
and the electric field is $E^{\mu }=g^{\mu \rho }\exp \left[ -4\alpha \phi
/(n-1)\right] F_{\rho \nu }u^{\nu }$. Then the electric charge per unit
volume $V_{n-1}$ can be found by calculating the flux of the electric field
at infinity, yielding
\begin{equation}
{Q}=\frac{\sqrt{\Xi ^{2}-1}q}{4\pi l^{n-2}}.
\end{equation}
Note that the electric charge of the system per unit volume is proportional
to the magnitude of rotation parameters or boost parameters, and is zero for
the case of a static solution. This result is expected since now, besides
the magnetic field along the $\phi ^{i}$ ($x^{i}$, for the case of angular
magnetic field) coordinates, there is also a radial electric field ($%
F_{tr}\neq 0$).\ To give a physical interpretation for the
appearance of the net electric charge, we first consider the
static spacetime. The magnetic field source can be interpreted as
composed of equal positive and negative charge densities, where
one of the charge density is at rest and the other one is spinning
(travelling, in the case of angular magnetic field). Clearly, this
system produce no electric field since the net electric charge
density is zero, and the magnetic field is produced by the
rotating (travelling) electric charge density. Now, we consider
the rotating (travelling) solutions. From the point of view of an
observer at rest relative to the source ($S$), the two charge
densities are equal, while from the point of view of an observe
$S^{\prime }$ that follows the intrinsic roation (translation) of
the spacetime, the positive and negative charge densities are not
equal, and therefore the net electric charge of the spacetime is
not zero.

\section{Concluding Remarks}

The counterterm method inspired by the AdS/CFT correspondence has been
widely used for the computation of the finite action and conserved
quantities of asymptotically AdS solutions of Einstein gravity. Testing the
validity of the counterterm method for nonasymptotically AdS spacetimes
needs new solutions which are not asymptotically AdS. In this paper, we
obtained two new classes of $(n+1)$-dimensional exact magnetic rotating
solutions of Einstein-Maxwell dilaton gravity in the presence of
Liouville-type potential, which are neither asymptotically flat nor AdS.
These solutions which are ill-defined for $\alpha =\sqrt{n}$ reduce to the
horizonless rotating solutions of \cite{Deh2} for $\alpha =0$, while for $%
n=3 $, they reduce to the four-dimensional magnetic dilaton string presented
in \cite{Deh4}. The first class of solutions represent a $(n+1)$-dimensional
spacetime with a longitudinal magnetic field. We found that these solutions
have no curvature singularities and no horizons, but have conic singularity
at $r=0$. We also confirmed that these solutions are both null and timelike
geodesically complete. In fact, we showed that every null or timelike
geodesic starting from an arbitrary point can either extend to infinite
values of the affine parameter along the geodesic or end on a singularity at
$r=0$. In these spacetimes, when all the rotation parameter are zero (static
case), the electric field vanishes, and therefore the brane has no net
electric charge. For the spinning brane, when one or more rotation
parameters are nonzero, the brane has a net electric charge density which is
proportional to the magnitude of rotation parameter given by $%
\sum_{i}^{k}a_{i}^{2}$. The second class of solutions represent a spacetime
with angular magnetic field. These solutions have no curvature singularity,
no horizon, and no conic singularity. Although a boost transformation gives
a solution with linear momentum which generates an electric field, it is not
a new solution. This is due to the fact that the boost transformation is
permitted globally. We also showed that, for the case of traveling brane
with one or more nonzero boost parameters, the net electric charge of the
brane is proportional to the magnitude of the velocity of the brane ($%
\sum_{i}^{\kappa }v_{i}^{2}$). Finally, we used the counterterm method and
calculated the conserved quantities of the two classes of solutions which
are nonasymptotically AdS. These calculations show that one may use the
counterterm method for calculating the conserved quantities of these
nonasymptotically AdS spacetimes.

\acknowledgments{This work has been supported in part by Research
Institute for Astrophysics and Astronomy of Maragha, Iran and also
by Shiraz University.}

\end{document}